\documentstyle[aps,prl,twocolumn,epsfig]{revtex}

\newcommand{\be}{\begin{eqnarray}}
\newcommand{\ee}{\end{eqnarray}}

\begin{document}

\twocolumn[\hsize\textwidth\columnwidth\hsize\csname @twocolumnfalse\endcsname

\title{Carbon substitution effect in MgB$_2$} 
\author{Jai Seok Ahn$^{1,\dag}$ and Eun Jip Choi$^{1,2,\ddag}$}
\address{$^1$Center for Strongly Correlated Materials Research, Seoul National University, Seoul 151-742, Korea}
\address{$^2$Department of Physics, University of Seoul, Seoul 130-743, Korea}
\date{\today }
\maketitle

\begin{abstract}
We investigated carbon substitution effect on boron plane of superconducting MgB$_2$. 
MgB$_2$ and MgB$_{1.8}$C$_{0.2}$ samples are synthesized under high pressure furnace.
MgB$_{1.8}$C$_{0.2}$ are characterized as AlB$_2$-type single phase with smaller 
B-B distance. During the superconducting transition, two distinct onset temperatures are 
observed in MgB$_{1.8}$C$_{0.2}$.

\end{abstract}

\pacs{PACS number: }

\vskip2pc]


Since the discovery of superconducting property in MgB$_{2}$ by Akimitsu 
{\it et al.}\cite{Akimitsu2001}, much efforts are spent on understanding the
type of carriers and the way of optimization with chemical substitution. As
far as we know, chemical substitution of elements are focused mostly on
Mg-site surrounded by neighboring 12 boron cages, with an exception of $^{10}
$B isotope effect by Bud'ko {\it et al}.\cite{Budko2001} Several alkali and
alkali earth and group III elements are thoroughly tested, such as Al\cite
{Slusky2001}, Be \cite{Felner2001}, and Li\cite{Zhao2001}, however, decrease
of transition temperature is observed in all cases, partly due to the
structural instability introduced by ion size difference between Mg and
substituting elements. If the structural uniqueness is essential in finding
superconducting borides, possibility may be low to find superconductivity
among the existing diboride compounds containing heavy elements as suggested
by structural analysis\cite{Bianconi2001}.

With these in mind, we focused on the chemical substitution on boron plane
itself, which carries superconducting property. On the crystallographic
point of view, boron shares similarity in structure with B$_{4}$C,
C(graphite), and BN in its rhombohedral phase and may be substituted with
those similar materials with high melting points. In this paper, we will
report our result on the carbon substitution effect on boron site of MgB$%
_{2} $.

MgB$_{2}$ and MgB$_{1.8}$C$_{0.2}$ samples are synthesized with high
pressure furnace.\ Starting materials are powders of Mg (99.8\%, Alfa),
amorphous B (99.99\%, Alfa), and carbon (high purity activated charcoal,
Wako). Stoichiometric amounts of powders are weighed, thoroughly mixed in an
agate mortar, and pressed into pellets. The pellets are placed in a tungsten
vessel with a close-fitting cap, reacted one hour at 850 $^{\circ }$C, and
subsequently annealed at 500 $^{\circ }$C for 12 hours under 20 atm of high
purity argon atmosphere. Some amounts of magnesium turnings (99.9\%, Acros)
are used as getters inside the furnace.

Figure 1(a) show x-ray $\theta $-2$\theta $ diffraction patterns of MgB$_{2}$
and MgB$_{1.8}$C$_{0.2}$ measured with Cu K$_{\alpha }$ source with 2$\theta 
$ step of 0.05$^{\circ }$. In both patterns, single group of AlB$_{2}$-type
reflections and small amounts of MgO impurities (shown with asterisks) are
observed and any other impurities of B, C, or Mg is not observed within
accuracy. Our calculation of hexagonal cell parameters for MgB$_{2}$, $%
a_{H}=3.083$ \r{A} and $c_{H}=3.520$ \r{A} is quite close to reported one: $%
3.086$ and $3.524$ \r{A}.(Here, peak positions of MgO are used as internal
standards.) Parameters are also calculated for MgB$_{1.8}$C$_{0.2}$: $%
a_{H}=3.070$ \r{A} and $c_{H}=3.520$ \r{A}. From these results, we notice
that only $a_{H}$(B-B distance) decreases and $c_{H}$(separation between B
planes) is not varied with carbon substitution. This is more clear from Fig.
1(b), showing the movement of (100), (101), and (002) lines with carbon
substitution.

Comparison with the result of other elements is intriguing at this point. In
Al$^{3+}$-substitution (hole doping), both $a$- and $c$-axis length decrease
with doping, and the change in $c$-axis is significant at high doping limits
($\geq $ 0.25 Al)\cite{Slusky2001}. For Be$^{2+}$-substitution (no doping,
ion size effect), it is well-known that Be cannot dope Mg site because it
forms stable phase of BeB$_{2}$\cite{Hoenig1961,Felner2001}. And in Li$^{+}$%
-substitution (electron doping), $c$-axis does not vary and $a$-axis length
decreases with doping concentration ($\leq $ 0.3 Li)\cite{Zhao2001}, quite
similar to our result: C$^{4+}$ ion is smaller than B$^{3+}$\cite
{Shannon1976} and electron doping is expected from ionic picture.

In order to see the changes in superconducting properties, dc resistivity is
measured on a closed-cycle He refrigerator (Janis, CCS350), with a standard
four-point probe technique with polished rectangular sample of dimension $%
\sim $ 2$\times $5$\times $1 mm$^{3}$, where electrical contact is made with
silver epoxy (Dotite) and thermal one with grease (Apiezon, N). Figure 2
shows the sample resistivities normalized with the values at 50 K and inset
shows the region of phase transition. Normal state resistivity level of MgB$%
_{2}$ and MgB$_{1.8}$C$_{0.2}$ is different by two orders of magnitude, such
as $\rho (50$K$)$ = 93 $\mu \Omega $cm (MgB$_{2}$) and 19 m$\Omega $cm (MgB$%
_{1.8}$C$_{0.2}$), partly due to the low density of our MgB$_{1.8}$C$_{0.2}$
pellet. However, their normal state power-law dependence($\rho \sim $ a $+$
bT$^{2}$) and residual resistivity ratio ($\rho $(T$_{\text{C}}^{+}$ )/$\rho 
$(300K) $\sim $ 0.55) are quite similar. In MgB$_{2}$, the resistivity drops
at 38.8 K and the transition is completed within 1 K, as shown in the inset,
which is quite close to the report of Akimitsu et al.\cite{Akimitsu2001}

Compared with MgB$_{2}$, the transition width is very broad($\sim $ 8 K) in
MgB$_{1.8}$C$_{0.2}$, which begins at 41 K and ends at 33 K. Also, the
resistivity shows two distinct onset temperatures indicated with $\nabla $
symbols in the inset (T$_{c1}\sim $ 41 K and T$_{c2}\sim $ 37 K).
Considering the single group of AlB$_{2}$-type patterns of x-ray, structural
inhomogeneity may not be related with this observation, however, some
intergrain effect should be counted. According to our preliminary results of
different doping concentrations, structural segregation begins at some
higher doping and MgB$_{1.8}$C$_{0.2}$ is near optimum in its property.
Because carbon substitution gives influences in effective mass and charge
valency of undoped boron planes, we speculate some changes in phonon or
electronic structure corresponding to this carbon substitution. In this
respect, further progresses with boron plane substitution are anticipated.

This work was supported by KRF-99-041-D00185 and by the KOSEF through the
CSCMR. We acknowledge the help of J. H. Bae, Y. J. Kim, and S. C. Kim of
Univ. of Seoul.%

\begin{figure}[tbp]
\caption{Dc resistivities nomalized at 50 K ($\protect\rho $(T)/$\protect%
\rho $(50K)) of MgB$_{2}$ and MgB$_{1.8}$C$_{0.2}$. Inset shows the region
of the transition temperatures.}
\label{Fig:2}
\end{figure}


\begin{references}
\bibitem[\dag]{Footnote1}  jsahn@physics.uos.ac.kr

\bibitem[\ddag]{Footnote2}  echoi@uoscc.uos.ac.kr

\bibitem{Akimitsu2001}  J. Nagamatsu, N. Nakagawa, T. Muranaka, Y. Zenitani,
and J. Akimitsu, Nature {\bf 410}, 63 (2001).

\bibitem{Budko2001}  S. L. Bud'ko, G. Lapertot, C. Petrovic, C. E.
Cunningham, N. Anderson, and P. C. Canfield, cond-mat/0101463 (2001).

\bibitem{Slusky2001}  J. S. Slusky, N. Rogado, K. A. Regan, M. A. Hayward,
P. Khalifah, T. He, K. Inumaru, S. Loureiro, M. K. Haas, H. W. Zandbergen
and R. J. Cava, cond-mat/0102262 (2001).

\bibitem{Felner2001}  Israel Felner, cond-mat/0102508 (2001).

\bibitem{Zhao2001}  Y. G. Zhao, X. P. Zhang, P. T. Qiao, H. T. Zhang, S. L.
Jia, B. S. Cao M. H. Zhu, Z. H. Han, X. L. Wang, and B. L. Gu,
cond-mat/0103077 (2001).

\bibitem{Bianconi2001}  A. Bianconi, N. L. Saini, D. Di Castro, S.
Agrestini, G. Campi, A. Saconne, S. De. Negri, M. Giovanni, and M.
Colapietro, cond-mat/0102410 (2001).

\bibitem{Hoenig1961}  C. L. Hoenig, C. F. Cline, and D. E. Sands, J. Am.
Cer. Soc. {\bf 44}, 385 (1961).

\bibitem{Shannon1976}  R. D. Shannon, Acta Cryst. A {\bf 32}, 751 (1976). 
\begin{figure}[tbph]
\caption{(a) X-ray $\protect\theta $-$2\protect\theta $ diffraction
patterns(Cu K$_{\protect\alpha }$) of MgB$_{2}$ and MgB$_{1.8}$C$_{0.2}$
powders. (b) Enlarged patterns in the region of the (100), (101), and (002)
reflections. MgO impurities are shown with asterisks.}
\label{Fig:1}
\end{figure}
\end{references}
\end{document}